\documentclass[a4paper,11pt]{article}
\usepackage{pos}

\usepackage{tikz}

\usepackage{enumitem}
\usepackage[T1]{fontenc} 
\usepackage{floatrow}
\usepackage{appendix}
\usepackage{slashed}
\usepackage{tabularx}
\usepackage[normalem]{ulem}

\allowdisplaybreaks

\usepackage{epsf}
\usepackage{empheq}
\usepackage[theorems,skins]{tcolorbox}

\providecommand{\U}[1]{\protect\rule{.1in}{.1in}}

\def\r{{\mathbf r}}

\def\slashchar#1{\setbox0=\hbox{$#1$}
   \dimen0=\wd0
   \setbox1=\hbox{/} \dimen1=\wd1
   \ifdim\dimen0>\dimen1
      \rlap{\hbox to \dimen0{\hfil/\hfil}}
      #1
   \else
      \rlap{\hbox to \dimen1{\hfil$#1$\hfil}}
      /
   \fi}

\def\bei{\begin{itemize}}
\def\ei{\end{itemize}}

\def\beeq{\begin{eqnarray}} 
\def\beqa{\begin{eqnarray}}
\def\bea{\begin{eqnarray}}

\def\eea{\end{eqnarray}}
\def\eqa{\end{eqnarray}}
\def\eeeq{\end{eqnarray}}

\def\eqar{\end{array}}
\def\beqar{\begin{array}}

\def\beas{\begin{eqnarray*}}
\def\beqas{\begin{eqnarray*}}

\def\eqas{\end{eqnarray*}}
\def\eeas{\end{eqnarray*}}

\def\beq{\begin{equation}} 
\def\be{\begin{equation}}

\def\ee{\end{equation}}
\def\eq{\end{equation}}
\def\eeq{\end{equation}}

\def\beqd{\begin{displaymath}}
\def\eeqd{\end{displaymath}}
\def\eqd{\end{displaymath}}

\def\beeq{\begin{eqnarray}} \def\eeeq{\end{eqnarray}}

\newcommand{\fin}{\end{document}}

\title{NLO corrections to the NEik DIS structure functions}

\author{Tolga Altinoluk}
\author{Guillaume Beuf}
\author*{Jules Favrel} 
\author{Michael Fucilla}

\affiliation[a]{National Centre for Nuclear Research (NCBJ), Pasteura 7, 02-093 Warsaw, Poland}

\emailAdd{tolga.altinoluk@ncbj.gov.pl}
\emailAdd{guillaume.beuf@ncbj.gov.pl}
\emailAdd{jules.favrel@ncbj.gov.pl}
\emailAdd{michael.fucilla@ncbj.gov.pl}

\abstract{
We summarize the next-to-leading order (NLO) corrections to next-to-eikonal (NEik) quark background contributions to DIS structure functions. At NEik accuracy, in addition to  corrections arising from the gluon background field of the target, DIS structure functions receive contributions from the $t$-channel quark exchanges, represented by insertions of the quark background field of the target. The latter provide the lowest order contributions in $\alpha_s$ at NEik accuracy. We show that the NLO corrections to the longitudinal NEik structure functions are finite, whereas those to the transverse NEik  structure functions exhibit rapidity and ultraviolet (UV) divergences. We analyze these divergences and extract the finite contributions.     
}

\begin{document} 
\maketitle
\flushbottom
\section{Introduction}
\label{sec:intro}
Deep inelastic scattering (DIS) on dense targets
provides a particularly clean probe of saturation dynamics. In this process, a virtual photon emitted by the incoming lepton splits into a quark–antiquark pair that subsequently scatters off the target. The multiple interactions between incoming partons and the dense target are treated within the Color Glass Condensate (CGC) formalism (see Refs. \cite{Gelis:2010nm} for review).

The forthcoming Electron–Ion Collider (EIC) in the USA is expected to provide a clean experimental environment with high luminosity for future DIS measurements. In order to fully exploit the future EIC data to investigate gluon saturation, precise theoretical predictions for observables within the CGC framework are needed, including NLO and/or NEik corrections. 

The central approximation in the CGC framework is the eikonal approximation, which amounts to retaining only the leading-power contributions in the high-energy limit and neglecting energy-suppressed terms. The first power-suppressed terms in this expansion are referred to as next-to-eikonal (NEik) corrections and can arise from both the gluon and quark background fields of the target. The NEik corrections to various observables arising from the gluon background field have been studied extensively (see, for example, Ref.~\cite{Altinoluk:2022jkk}). In this work, we focus primarily on the NEik quark-background contributions to DIS structure functions and compute their next-to-leading (NLO) corrections in $\alpha_s$, while also considering the corresponding contributions arising from the gluon background field.

This contribution is organized as follows. In Section~\ref{Sec:Set_up}, we introduce the setup by reviewing the DIS cross section and the leading-order NEik DIS structure functions derived in Ref.~\cite{Altinoluk:2025ang} for longitudinally and transversely polarized photons. In Sections~\ref{Sec:Quark-Bckg} and \ref{Sec:Gluon-Bckg}, we present, respectively, the NLO corrections to the NEik DIS structure functions arising from the quark and gluon backgrounds. This analysis is  originally performed in Ref.~\cite{Altinoluk:2025ivn} and constitutes a first step toward a complete description of observables computed simultaneously at NLO and NEik accuracy within the CGC framework.

\section{Set up and definitions}
\label{Sec:Set_up}
In the one-photon exchange approximation, after integration over the azimuthal angle of the scattered lepton, the unpolarized DIS inclusive cross section can be written as, 
\begin{gather}
    \frac{d\sigma^{l+A\rightarrow l'+A}}{dx_{Bj}dQ^2} = \frac{\alpha_\text{em}}{\pi x_{Bj}Q^2}\bigg[\left(1-y+\frac{y^2}{2}\right)\sigma^{\gamma^*}_{T}\left(x_{Bj},Q^2\right)+(1-y)\sigma^{\gamma^*}_{L}\left(x_{Bj},Q^2\right)\bigg]\;,
\end{gather}
where $Q^2$ is the virtuality of the incoming photon, $x_{Bj}$ is the Bjorken scaling variable and $y$ is the inelasticity of the process. Moreover, $\sigma^{\gamma^*}_{L,T}$ correspond to total cross sections for the scattering of longitudinal or transverse photon and they are related to the structure functions via   
\begin{align}
\label{sigma2F}
\sigma^{\gamma^*}_{T,L}(x_{Bj},Q^2)=\frac{(2\pi)^2\, \alpha_{\text{em}}}{Q^2} F_{T,L}(x_{Bj},Q^2) \; .
\end{align}
The transverse or longitudinal structure functions at NEik accuracy was computed recently in \cite{Altinoluk:2025ang}, where the analysis focused exclusively on NEik corrections arising from $t$-channel quark exchanges. It was shown that quark background field contributions to DIS structure functions start at ${\cal O}({\alpha_s}^0)$ and can be expressed in terms of quark and antiquark parton distribution functions (PDFs) as 
\begin{align}
\label{FT_LO_fin}
F_T(x_{Bj},Q^2)\Big|_{{\rm LO}, \Psi+\overline{\Psi}}&=\sum_{f} e_f^2\, x_{Bj}\Big[ q_f(x_{Bj})+\overline{q}_f(x_{Bj})\Big]+{\rm NNEik} \, , \\
F_L(x_{Bj},Q^2)\Big|_{{\rm LO}, \Psi+\overline{\Psi}}&= 0 +{\rm NNEik}  \, , 
\end{align}
where the background quark PDF is defined by,
\begin{align}
    q_f\left(x_{Bj}\right) = \int \frac{d z^+}{2 \pi} e^{-ix_{Bj} p_t^- z^+} \langle p_t| \overline{\Psi} (z^+, \mathbf{0}) \mathcal{U}_F (z^+, 0^+, \mathbf{0}) \frac{\gamma^{-}}{2} \Psi(0^+, \mathbf{0}) | p_t \rangle \; .
    \label{Eq:QuarkPDFDef}
\end{align}
The background anti-quark PDF is deduced from the background quark PDF thanks to the relation $\overline{q}_f\left(x_{Bj}\right)=-q_f\left(-x_{Bj}\right)$. The NLO corrections to the NEik longitudinal and transverse DIS cross sections can be organized as 
\begin{gather}
    \sigma^{\rm L/T}_{{\rm NLO+NEik}}= \sigma^{\rm L/T}_{{\rm NLO},\Psi+\overline{\Psi}} +\sigma^{\rm L/T}_{{\rm NLO},\rm G}+\sigma^{\rm L/T,\;b.s}_{{\rm NLO},\rm G}\;.\label{Eq.FullXS}
\end{gather}
Here, the first term on the r.h.s correspond to the NLO corrections to the NEik quark (and antiquark) background and the second term to the NEik+NLO correction from the pure gluon background in the static approximation. The third term corresponds to NEik corrections  beyond the static (b.s.) approximation in gluon background which is shown to vanish at NEik order \cite{Altinoluk:2025ivn}. In the rest of this proceedings, we present the first two terms on the r.h.s. of Eq. \eqref{Eq.FullXS} for massless quarks. The computation of the massive case with full details can be found in \cite{Altinoluk:2025ivn}. 
\section{NEik DIS cross section at NLO : Fermion background contribution}
\label{Sec:Quark-Bckg}
\begin{figure}[h!]
    \centering
    \includegraphics[width=0.47 \linewidth]{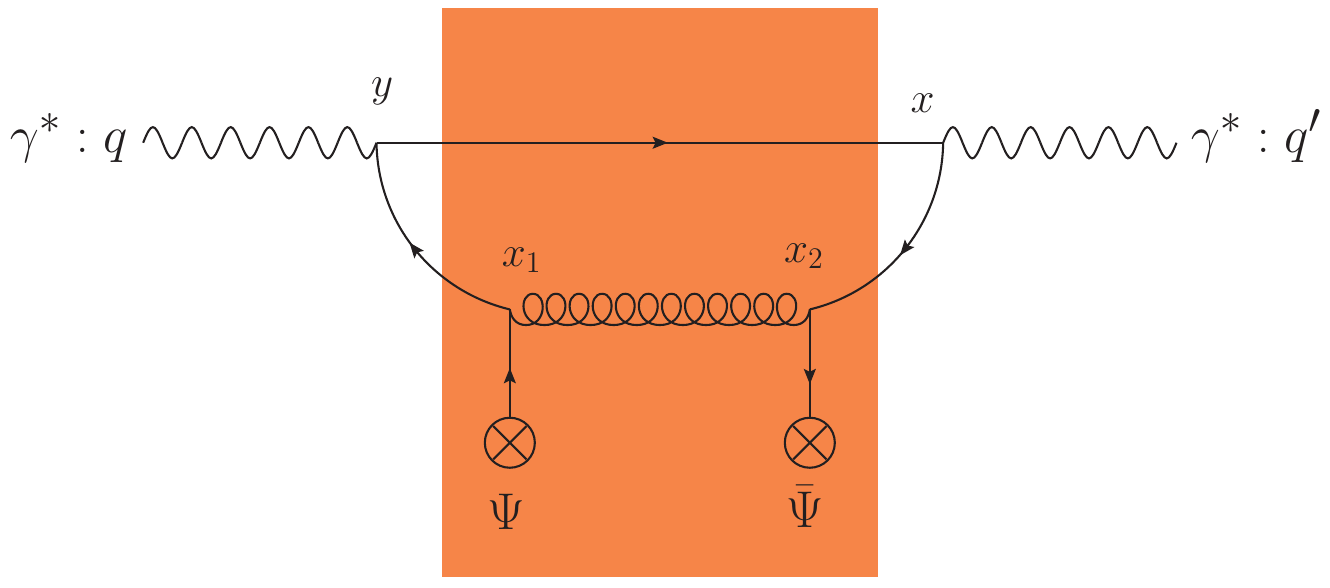} \hspace{0.5 cm}
    \includegraphics[width=0.47 \linewidth]{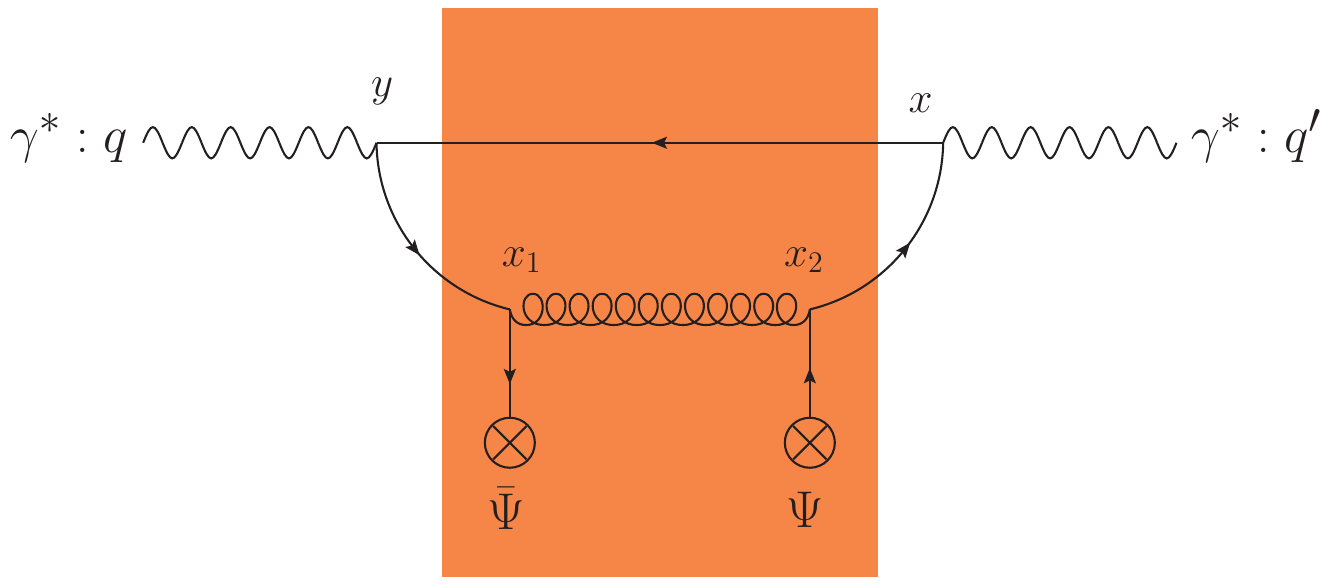}
    (a) \hspace{7.2 cm} (b)
    \caption{(a) NLO contribution to DIS at the NEik accuracy from the quark background of the target. (b) NLO contribution to DIS at the NEik accuracy from the antiquark background of the target.}
    \label{fig:NLODISNEik}
\end{figure}
\paragraph*{Longitudinal NEik DIS cross section at NLO.}
The NLO correction to the NEik fermion background contribution the to DIS longitudinal cross section reads \cite{Altinoluk:2025ivn} 
\begin{equation}
    \sigma_{{\rm NLO}, \Psi+\overline{\Psi}}^{\rm L}  = 8\, e_f^2\, \alpha_{\text{em}}\, \alpha_{s}\, x_{Bj}
    \int d^{2} \mathbf{r} \int_0^1 dz \; z^2 (1-z) K_{0}^2 (\bar{Q} |\mathbf{r}|) \; {\rm Re} : f\left(\mathbf{r},x_{Bj}\right)\; ,\label{Eq:Sigma_NLO_Psi_L}
\end{equation} 
where $\bar{Q}=\sqrt{z(1-z)Q^2}$ in the massless limit, $z$ corresponds to the longitudinal momentum fraction of the exchanged gluon with the photon. The relevant quark propagators in the mixed quark and gluon backgrounds can be found in \cite{Altinoluk:2024dba,Altinoluk:2025ivn}. The fermion background operator is defined as the sum of quark and anti-quark operators 
\begin{equation}
    f\left(\mathbf{r},x_{Bj}\right) = f_q\left(\mathbf{r},x_{Bj}\right)+f_{\bar{q}}\left(\mathbf{r},x_{Bj}\right)\;,\label{eq:def_F_main}
\end{equation}
that are given by
\begin{align}
\label{eq:def_F_q_main}
f_q\left(\mathbf{r},x_{Bj}\right) & =  \int \frac{d z^+}{2 \pi}\;e^{-ix_{Bj}p^-_tz^+} \, \theta (z^+)\; \big\langle p_t\big| \overline{\Psi} (z^+, \mathbf{0}) \gamma^- t^a \mathcal{U}_F\left( z^+,  \infty^+, \mathbf{0} \right) \mathcal{U}_F\left(\mathbf{r}  \right) 
    \nonumber \\ 
    & \hspace{3.5cm}
    \times \mathcal{U}_F \left( - \infty^+ , 0^+ , \mathbf{0} \right)t^b \Psi (0^+ , \mathbf{0}) \left[ \mathcal{U}_A \left(0^{+}, z^{+}, \mathbf{0} \right) \right]^{ba}  \big| p_t \big\rangle \; ;
\end{align}
\begin{align}
\label{eq:def_F_qbar_main}
f_{\bar q}\left(\mathbf{r},x_{Bj}\right)
&=  
\int \frac{d z^+}{2 \pi} \; e^{-ix_{Bj}p^-_tz^+} \,\theta (z^+)
\big\langle p_t\big| {\rm Tr_{D,c}} 
\Big[ 
\mathcal{U}_F \left( \infty^+ , z^+ , \mathbf{0} \right) t^a \Psi (z^+ , \mathbf{0}) 
\nonumber &\hspace{0.7cm}\\
& \hspace{1.2cm}
\times 
\left[ \mathcal{U}_A \left(z^{+}, 0^{+}, \mathbf{0} \right) \right]^{ab} \overline{\Psi} (0^+, \mathbf{0})   t^b \gamma^- \mathcal{U}_F\left( 0^+, - \infty^+, \mathbf{0} \right) \mathcal{U}_F^{\dagger} \left(\mathbf{r}  \right)  \Big] \big| p_t \big\rangle \; .
\end{align}
respectively. 
Due to the explicit factor of $x_{Bj}$ multiplying ${\rm Re}:f\left(x_{Bj},\r\right)$ in Eq. \eqref{Eq:Sigma_NLO_Psi_L}, the contribution from the $x_{Bj}$-dependent phase factor in the background operator defined in Eq. \eqref{eq:def_F_main} is of NNEik order.
At zero transverse separation, the real part of the fermion background operator reduces to the singlet quark background operator:
%
\begin{equation}
    {\rm Re}:f\left(x_{Bj},\mathbf{0}\right) = C_F\,\big[q_f\left(x_{Bj}\right) +\overline{q}_f\left(x_{Bj}\right)\big]\;.
\end{equation}
The NLO correction constitutes the first nonvanishing NEik quark-background contribution to the longitudinal DIS cross section. This contribution is free of both UV and rapidity divergences. Using the relation in Eq.~\eqref{sigma2F} together with Eq.~\eqref{Eq:Sigma_NLO_Psi_L}, we obtain the NLO correction to the NEik fermion background contribution to the longitudinal DIS structure function as 
\begin{align}
    F_L(x_{Bj},Q^2)\Big|_{{\rm NLO}, \Psi+\overline{\Psi}}&= 4\, x_{Bj} Q^2 \sum_{f} e_f^2 \, \, \frac{\alpha_{s}}{2 \pi} \, 
  \nonumber \\ & \times \int \frac{d^{2} \mathbf{r}}{\pi} \int_0^1 dz \; z^2 (1-z) K_{0}^2 (\bar{Q} |\mathbf{r}|) \; {\rm Re} : f(x_{Bj},\mathbf{r}) +{\rm NNEik}  \, .
\end{align}
\paragraph*{Transverse NEik DIS cross section at NLO.} The NEik fermion background contribution to the transverse DIS cross section at NLO reads \cite{Altinoluk:2025ivn} 
\begin{align}
    \sigma_{{\rm NLO}, \Psi+\overline{\Psi}}^{\rm T}  & 
    =  \frac{e_f^2 g^2 g_s^2  }{2 } \frac{x_{Bj}}{Q^2}  {\rm Re} : 
    \int_0^1 \frac{dz}{z}  \int \frac{ d^{2-2 \epsilon} \mathbf{r} }{(2 \pi)^{2-2 \epsilon}} 
    \frac{1}{(\mathbf{r}^2)^{-\epsilon}} 
    \Big[ \big[z^2 + (1-z)^2\big] - \epsilon \;  \Big]
    \nonumber \\
    & \hspace{7cm}
    \times \, 
    \big(\bar{Q}^2\big)^{1-\epsilon}\,  K_{1-\epsilon}^2 \big(\bar{Q} | \mathbf{r} |\big) \,  f (x_{Bj}, \mathbf{r})\;
    \label{Eq:Sigma_NLO_Psi_T_Massless}
\end{align}
with the quark background operator defined in Eq. \eqref{eq:def_F_main} together with Eqs. \eqref{eq:def_F_q_main} and \eqref{eq:def_F_qbar_main}. The NLO correction to the transverse NEik DIS cross section, given in Eq. \eqref{Eq:Sigma_NLO_Psi_T_Massless}, exhibits a rapidity divergence in the $z\rightarrow0$ limit for non vanishing transverse separation $\r$. Since the longitudinal phase space of the emitted gluon extends to arbitrarily small $z$, a large rapidity separation arises between the hard parton traversing the entire target (the quark in diagram (a) or the antiquark in diagram (b) of Fig.~\ref{fig:NLODISNEik}) and the parton emitting the gluon (the antiquark in diagram (a) or the quark in diagram (b)). As shown in Eq.~\eqref{FT_LO_fin}, the LO contribution to the inclusive DIS cross section for a transversely polarized photon is proportional to the background quark distribution. This LO quark-background contribution is purely NEik and therefore does not mix with the eikonal evolution. The expression in Eq.~\eqref{Eq:Sigma_NLO_Psi_T_Massless} represents the corresponding NLO correction. Consequently, the contribution arising in the $z\rightarrow 0$ limit should be interpreted as part of the small-$x$ evolution of the singlet quark background distribution, $q_f(x)+\overline{q}_f(x)$.
%

In addition to the rapidity divergences discussed above, the NLO correction to the NEik quark background contribution to the inclusive DIS cross section for a transversely polarized photon, given in Eq. \eqref{Eq:Sigma_NLO_Psi_T_Massless}, also exhibits
UV divergences arising in the limit $\r \rightarrow \mathbf{0}$ for $0 < z < 1$. The rapidity and UV divergences overlap in the simultaneous $z\rightarrow 0$ and $\r\rightarrow\mathbf{0}$ limit, suggesting the emergence of double-logarithmic evolution.

%
The UV divergences are regulated using the standard dimensional regularization, while the rapidity divergence in Eq. \eqref{Eq:Sigma_NLO_Psi_T_Massless} is regulated using the $\eta^+$ regulator, see Ref. \cite{Altinoluk:2025tms}. Accordingly, the $z$-integration in Eq. \eqref{Eq:Sigma_NLO_Psi_T_Massless} is replaced by
\begin{align}
    \int_0^1 \frac{dz}{z}\longrightarrow\int^1_0\frac{dz}{z}\left(\frac{zq^+}{\nu^+}\right)^{\eta}\;.
\end{align}
In the following, we use the relation $q^+=Q^2/(2x_{Bj}p^-_t)$. This allows us to decompose the NLO correction to the NEik transverse DIS structure function into finite and divergent contributions:
\begin{align}
    F_T\left(x_{Bj},Q^2;\{\mu_R^2,\epsilon\},\{\nu^+,\eta\}\right)\bigg|_{\rm NLO,\,\Psi+\overline{\Psi}}=F_T\left(x_{Bj},Q^2\right)\bigg|^{\rm finite}_{\rm NLO,\,\Psi+\overline{\Psi}}\hspace{3.0cm}\nonumber\\\hspace{6.4cm}+\;F_T\left(x_{Bj},Q^2;\{\mu_R^2,\epsilon\},\{\nu^+,\eta\}\right)\bigg|^{\rm div.}_{\rm NLO,\,\Psi+\overline{\Psi}}\;.\label{FT_NLO_Quarl}
\end{align}
The finite piece depends only on the DIS variables $(x_{Bj},Q^2)$, whereas the divergent part depends explicitly on the scale-regulator pairs $\{\mu^2_R,\epsilon\}$ for the UV divergence $\{\nu^+,\eta\}$ for the rapidity divergence. The UV renormalization scale $\mu^2_R$ is related to the dimensional-regularization scale  $\mu^2=\mu^2_Re^{\gamma_e}/(4\pi)$. As discussed above, three types of divergences arise at NLO: a pure UV divergence, a mixed UV and rapidity divergence, and a pure rapidity divergence. They are given, respectively, by
\begin{align}
    F_T\left(x_{Bj},Q^2;\{\mu_R^2,\epsilon\}\right)\bigg|^{\rm UV\;div.}_{\rm NLO,\,\Psi+\overline{\Psi}} = \frac{\alpha_sC_F}{2\pi}\bigg[-\frac{1}{\epsilon}+\log\Bigg(\frac{Q^2}{\mu^2_R}\Bigg)-1+O(\epsilon)\bigg]F_{T}\left(x_{Bj},Q^2\right)\bigg|_{\rm LO,\,\Psi+\overline{\Psi}}\;;
\end{align}
\begin{gather}
    F_T\left(x_{Bj},Q^2;\{\mu_R^2,\epsilon\},\{\nu^+,\eta\}\right)\bigg|^{\rm UV\,\&\, rap.\;div.}_{\rm NLO,\,\Psi+\overline{\Psi}} = \frac{\alpha_sC_F}{2\pi}\bigg[\frac{1}{\eta}+\log\Bigg(\frac{1}{x_{Bj}}\Bigg)+\log\Bigg(\frac{Q^2}{2\nu^+p^-_t}\Bigg)+O(\eta)\bigg]\nonumber\hspace{2.0cm}\\\hspace{5.1cm}\times\;\bigg[\frac{1}{\epsilon}+\log\Bigg(\frac{\mu^2_Rb^2_0}{4}\Bigg)-1+O(\epsilon)\bigg]F_{T}\left(x_{Bj},Q^2\right)\bigg|_{\rm LO,\,\Psi+\overline{\Psi}}\;;
\end{gather}
\begin{gather}
    F_T\left(x_{Bj},Q^2;\{\nu^+,\eta\}\right)\bigg|^{\rm rap.\;div.}_{\rm NLO,\,\Psi+\overline{\Psi}} = \sum_fe^2_f\,\frac{\alpha_sC_F}{2\pi}\bigg[\frac{1}{\eta}+\log\Bigg(\frac{1}{x_{Bj}}\Bigg)+\log\Bigg(\frac{Q^2}{2\nu^+p^-_t}\Bigg)+O(\eta)\bigg]\nonumber\hspace{2.0cm}\\\hspace{5.85cm}\times\;{\rm Re}:\int\frac{d^2\r}{\pi \,\r^2}\bigg(f\left(x_{Bj},\r\right)-\theta\left(b^2_0e^{-2\gamma_e}-\r^2\right)f\left(x_{Bj},\mathbf{0}\right)\bigg)\;.
\end{gather}
Here $\gamma_e$ is the Euler–Mascheroni constant. Finally, the finite NLO contribution to the NEik transverse structure function is found to be,
\begin{align}
    F_T(x_{Bj} ,Q^2)\bigg|^{\rm finite}_{{\rm NLO}, \Psi+\overline{\Psi}}
    =
    x_{Bj} \sum_{f}
    e_f^2 \frac{\alpha_s}{2 \pi}
    \,{\rm Re} :
    \int_0^1 \frac{dz}{z}
    \int \frac{ d^{2} \mathbf{r} }{ \pi }
    \bigg\{
  \Big[ \bar{Q}^2 K_{1}^2\big(\bar{Q}|\mathbf{r}|\big) - \frac{1}{\mathbf{r}^2} \Big]
        f(x_{Bj}, \mathbf{r}) 
        \nonumber \\\hspace{2.3cm}
        -\; 2 z(1-z)\,\bar{Q}^2 K_{1}^2\big(\bar{Q}|\mathbf{r}|\big)
        \Big[ f(x_{Bj}, \mathbf{r}) - f(x_{Bj}, \mathbf{0}) \Big]
    \bigg\}  \; .
\end{align} 
\section{NEik DIS cross section at NLO : Gluon background contribution}
\label{Sec:Gluon-Bckg}
\begin{figure}[h!]
    \centering
    \includegraphics[width=0.60 \linewidth]{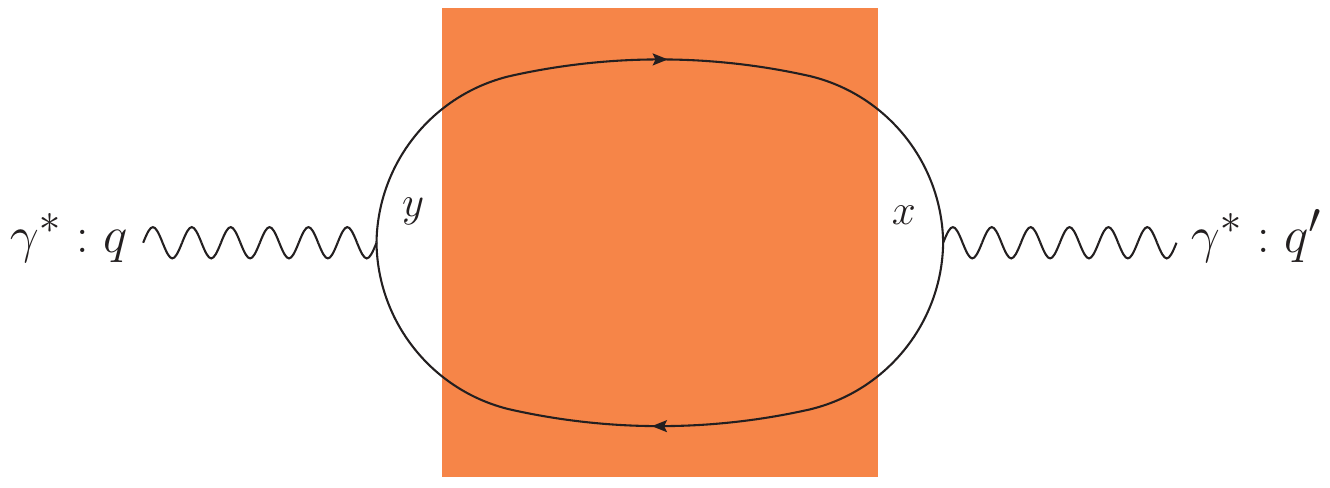} 
    \caption{NLO contribution to DIS at the NEik accuracy from the gluon background of the target.}
    \label{fig:NLODISNEikgluon}
\end{figure}
The diagram shown in Fig.\eqref{fig:NLODISNEikgluon} gives the LO contribution in dipole factorization, at both  eikonal and NEik order, in a pure gluon background field. The quark propagator in gluon background was computed at NEik accuracy in Ref. \cite{Altinoluk:2022jkk}. In the dense-target regime, the gluon background field scales as $A^- = O(1/g_s)$, and the diagram therefore yields a contribution of $O(1)$. However, additional care is required when NEik effects associated with quark background fields are included in the analysis of observables. The Yang-Mills equations imply that, in the dense regime, the quark background field $\Psi$ scales at most as $O(1/g_s)$.  Therefore, the LO DIS structure functions given in Eq. \eqref{FT_LO_fin} scales as $O(1/g_s^2)$. Consequently, in the dense-target regime, the NEik contribution in a gluon background field shown in Fig.\eqref{fig:NLODISNEikgluon} constitutes a relative $O(g_s^2)$ correction to the quark background contribution to the LO DIS structure functions.

Like the fermion background operator defined in Eq. \eqref{eq:def_F_main}, the gluon background operator can be decomposed as $G(x_{Bj},\r)=G_q(x_{Bj},\r)+G_{\bar{q}}(x_{Bj},\r)$ at NEik accuracy. The two terms are obtained by performing NEik expansion of the quark and anti-quark propagators, respectively, and are given by
\begin{align}
\label{de:Gq_2}
 G_q (x_{Bj},\mathbf{r}) &= - \frac{i}{g_s^2}  
 \int \frac{d z^{+}}{ \pi} \theta ( z^+ )  z^+  e^{-ix_{Bj}{{p_t^-}} z^+}\;  
  \big\langle p_t\big|  {\rm Tr_c}  \Big[  \mathcal{U}_{F} (\mathbf{r})   \mathcal{U}_F\left(-\infty^+, 0^+ ; \mathbf{0} \right) 
  \nonumber \\ 
  &
  \times  \; g_st \cdot \mathcal{F}_j^{\; -} (0^+, \mathbf{0}  ) \; \mathcal{U}_F\left( 0^{+}, z^+ ; \mathbf{0} \right) g_st \cdot \mathcal{F}_j^{\; -} (z^+ , \mathbf{0}  ) \mathcal{U}_F\left( z^+ , \infty^+ ; \mathbf{0} \right) \big] \big| p_t \big\rangle \;  ;
\end{align}
\begin{align}
    G_{\bar{q}} (x_{Bj},\mathbf{r}) & = -\frac{i}{g_s^2}  \int \frac{d z^{+}}{ \pi} \theta ( -z^+ )  (-z^+) \, e^{-ix_{Bj}p_t^- z^+}\;   \big\langle p_t\big|  {\rm Tr_c}  \Big[  \mathcal{U}^{\dagger}_{F} (\mathbf{r})  \,  \mathcal{U}_F \left( \infty^+, 0^+ ; \mathbf{0} \right) 
    \nonumber \\ 
  &  \hspace{1cm}
  \times  g_st \cdot \mathcal{F}_j^{\; -} (0^+, \mathbf{0}  ) \; \mathcal{U}_F\left( 0^{+}, z^+ ; \mathbf{0} \right) g_st \cdot \mathcal{F}_j^{\; -} (z^+ , \mathbf{0}  )\,  \mathcal{U}_F\left( z^+ , -\infty^+ ; \mathbf{0} \right) \Big ] \big| p_t \big\rangle \; .
\end{align}
At zero transverse separation the gluon background operator reduces to
\begin{align}
    {\rm Re}: G(x_{Bj},\mathbf{0})= \frac{\partial}{\partial {\rm x}} \Big[{\rm x} g ({\rm x})\Big] \bigg|_{{\rm x}=x_{Bj}}\;,
\end{align}
with the background gluon PDF defined as
\begin{align}
 g (x_{Bj}) = \frac{1}{x_{Bj} p_t^{-}} \int \frac{dz^+}{2\pi} e^{-ix_{Bj}p_t^{-}z^+ }   \langle p_t| \mathcal{F}_j^{\; - a} (0^+, \mathbf{0}  ) \; [ \mathcal{U}_A \left( 0^{+}, z^+ ; \mathbf{0} \right)]^{ab} \mathcal{F}_j^{\; - b} (z^+ , \mathbf{0}  )   | p_t \rangle \; .
\end{align}
The only difference between the NLO corrections arising from the gluon and quark background fields at NEik accuracy lies at the operator level. Therefore, the NLO corrections to the NEik gluon background contribution to inclusive cross sections can be obtained through the replacement
\begin{equation}
\label{rel_F2G}
    f(x_{Bj},\mathbf{r}) \longrightarrow  \frac{2}{D-2} G(x_{Bj},\mathbf{r}) 
\end{equation}
in Eq. \eqref{Eq:Sigma_NLO_Psi_L} for a longitudinally polarized photon and in Eq. \eqref{Eq:Sigma_NLO_Psi_T_Massless} for transversely polarized photon.
\vspace{1.5mm}\\
\paragraph*{Conclusion and Outlook.} 
The NLO corrections to the NEik DIS structure function for a transversely polarized photon contain a contribution that suggests the emergence of a double-logarithmic evolution. We are currently working on the derivation of the corresponding evolution of the quark operator directly from its operator definition at low $x_{Bj}$. 
\acknowledgments
TA and JF are supported in part by the National Science Centre (Poland) under the research Grant No. 2023/50/E/ST2/00133 (SONATA BIS 13). GB is supported in part by the National Science Centre (Poland) under the research Grant No. 2020/38/E/ST2/00122 (SONATA BIS 10). The work of MF is supported by the ULAM fellowship program of NAWA No. BNI/ULM/2024/1/00065 “Color glass condensate effective theory beyond the eikonal approximation”.
\bibliographystyle{apsrev}
\bibliography{mybib_New}

\end{document}